\documentclass[12pt]{article}
\usepackage{amsmath,amssymb,mathrsfs,url}
\usepackage{cite}

\DeclareMathAlphabet{\mathpzc}{OT1}{pzc}{m}{it}

\title{Wheeler-DeWitt quantization for point-particles}

\author{Ward Struyve\footnote{Instituut voor Theoretische Fysica, KU Leuven, Belgium}$^{*}$\footnote{Centrum voor Logica en Filosofie van de Wetenschappen, KU Leuven, Belgium}  }

\def\de{\delta}

\def\na{\nabla}
\def\lam{\lambda}

\def\ka{\kappa}

\def\pa{\partial}

\def\ka{\kappa}
\def\ii{\textrm i}
\def\ee{\textrm e}

\def\bn{{\boldsymbol \nabla}}

\newcommand{\be}{\begin{equation}}
\newcommand{\en}{\end{equation}}
\newcommand{\bi}{\begin{itemize}}
\newcommand{\ei}{\end{itemize}}

\begin{document}
\maketitle

\begin{abstract}
  \noindent
We present the Hamiltonian formulation of a relativistic point-particle coupled to Einstein gravity and its canonical quantization \`a la Wheeler-DeWitt. In the resulting quantum theory, the wave functional is a function of the particle coordinates and the 3-metric. It satisfies a particular Hamiltonian and diffeomorphism constraint, together with a Klein-Gordon-type equation. As usual in the Wheeler-DeWitt theory, the wave function is time-independent. This is also reflected in the Klein-Gordon-type equation, where the time derivative is absent. Before considering gravity, we consider the coupling of a particle with electromagnetism, which is treated similarly, but simpler.
\end{abstract}

\section{Introduction}
Canonical quantization of Einstein gravity leads to the Wheeler-DeWitt theory where the state is described by a wave functional $\Psi(h_{ij})$, which is a function of a 3-metric $h_{ij}({\bf x})$ \cite{kiefer04}. A common way to introduce matter is through a scalar field, so that the wave functional also is a function of a scalar field $\phi({\bf x})$, i.e., $\Psi=\Psi(h_{ij},\phi)$. A remarkable feature is that the wave functional turns out to be time-independent. The wave equations are merely constraints on the wave functional: the Hamiltonian and the diffeomorphism constraint. The time-independence of $\Psi$ leads to the question of how time evolution can be accounted for. There are various attempts to answer this question, but there is far from a consensus on the right approach \cite{kuchar92,rovelli04,kiefer04,pinto-neto19}.

In this paper, we consider the coupling of gravity to point-particles instead of a scalar field. This allows for the study of quantum gravity for a fixed number of particles without having to consider quantum field theory. This system has been studied in the context of the problem of time by Rovelli \cite{rovelli91a,rovelli91b} and Pav{\v s}i{\v c} \cite{pavsic11,pavsic20}. 

Classically, the history of a point-particle is described by its world-line $X^\mu(\lam)$, which depends on an arbitrary parameter $\lam$, while the space-time metric  $g_{\mu \nu}(x)$ is a function of space-time $x=(t,{\bf x})$.{\footnote{Units are assumed such that $\hbar=c=1$.}} The particle and field are hence parameterized differently, with $X^0(\lam) = t$. To pass to the Hamiltonian picture, a common temporal parameter for the field and the particle needs to be chosen. This is already familiar from the case of the free particle in Minkowski space-time. In that case, the disadvantage of choosing the parameter $t$ is that one needs to take a square root to express the Hamiltonian in terms of the momenta, resulting in the Hamiltonian $H = \pm\sqrt{{\bf p}^2 + m^2}$. Choosing the parameter $\lam$ instead leads to a zero Hamiltonian, but with the constraint $p_\mu p^\mu + m^2 = 0$. Quantization of the latter leads to the Klein-Gordon equation, while the former leads to the positive or negative frequency part of the Klein-Gordon equation. 

The situation in the case of gravity is similar. Rovelli \cite{rovelli91a,rovelli91b} considered the parameter $t$ and ended up taking a square root. Pav{\v s}i{\v c} seems to use $t$ for the field, while using $\lam$ for the particle. This does not amount to a standard Hamiltonian formulation and hence makes this quantization of the theory questionable. In this paper, the Hamiltonian formulation is presented using $\lam$ as the common temporal parameter, together with the corresponding quantum theory. This avoids needing to take the square root and as such generalizes Rovelli's treatment. The wave function is again time-independent; It does not depend on $X^0$, just as in Rovelli's treatment, but contrary to the findings of Pav{\v s}i{\v c}. So $X^0$ does not appear as a possible candidate for the evolution parameter (contrary to for example the treatment of dust \cite{brown95}). It will appear however that Pav{\v s}i{\v c}'s theory (when suitably interpreted) can be derived from ours in a mixed Schr\"odinger-Heisenberg picture. 

To start with, the treatment of point-particles coupled to an electromagnetic field is considered, in Minkowski space-time, which shares many features with the gravitational case.

\section{Electrodynamics}\label{elec}
To formulate the quantum theory of a point-particle interacting with an electromagnetic field, we start from the classical theory, pass to the Hamiltonian picture, and apply the usual canonical quantization methods \cite{dirac64,hanson76,sundermeyer82,gitman90,henneaux91}. 

Consider a classical point-particle with worldline $X^\mu(\lam)$ interacting with an electromagnetic field with vector potential $A^\mu=(A_0,{\bf A})$. Derivatives with respect to $\lam$ and $t$ will be denoted by respectively by primes and dots throughout, e.g.\ $X'^\mu = dX^\mu/d\lam$. It is also assumed throughout that $X'^0 > 0$. The action is 
\be
{\mathcal S} =  {\mathcal S}_M + {\mathcal S}_F ,
\en
with
\be
{\mathcal S}_M = - m \int d\lambda \sqrt{{X'}^\mu(\lam) X'_\mu (\lam)} - e \int d\lambda {X'}^\mu(\lam) A_\mu(X(\lam))
\en
and
\be
{\mathcal S}_F = -\frac{1}{4} \int d^4x F_{\mu \nu}(x) F^{\mu \nu}(x),
\en
where $F^{\mu \nu}=\pa^\mu A^\nu - \pa^\nu A^\mu$ is the electromagnetic field tensor. The temporal parameter in the matter action and the field action are different. In the matter action, the parameter $\lambda$ acts as temporal parameter, while $t$ does it for the electromagnetic field. In order to pass to a Hamiltonian formulation, a common temporal coordinate is needed. We will first use $t$ and then $\lambda$. 

To write the matter action in terms of $t$, the identity $1= \int dt \delta(t - X^0(\lam))$ is inserted in the action and an integration over $\lambda$ is performed. Defining ${\widetilde X}^\mu(t) = X^\mu(\lam(t)) $, where $X^0(\lam(t))\equiv 1$, leads to
\be
{\mathcal S}_M = - m \int dt \sqrt{1 - \dot {\widetilde {\bf X}}(t)^2} + e \int dt \dot {\widetilde {\bf X}}(t) \cdot {\bf A}(t,{\widetilde {\bf X}}(t)) - e \int dt  A_0(t,{\widetilde {\bf X}}(t)).
\en
Writing ${\mathcal S} = \int dt L({\widetilde {\bf X}},\dot {\widetilde{\bf X}}, A, \dot A)$, with $L$ the Lagrangian, the canonical momenta are 
\be
{\widetilde {\bf P}} = \frac{\pa L}{\pa \dot {\widetilde {\bf X}}} = m \frac{{\widetilde {\bf X}} }{\sqrt{1 - \dot {\widetilde {\bf X}}^2}} +e{\bf A}({\widetilde {\bf X}}) , 
\label{pmom}
\en
\be
\Pi_0({\bf x}) = \frac{\de L}{\de \dot A_0({\bf x})} = 0, \qquad {\boldsymbol \Pi}({\bf x}) = \frac{\de L}{\de \dot {\bf A}({\bf x})} = \dot {\bf A}({\bf x}) + \bn A_0({\bf x}). 
\label{fmom}
\en
The canonical Hamiltonian is 
\begin{align}
H &= {\bf P}\cdot \dot {\widetilde {\bf X}} + \int d^3 x( \pi_0 \dot A_0 + {\boldsymbol \Pi} \cdot  \dot {\bf A}) - L \nonumber\\
&= \frac{m}{ \sqrt{1 - \dot {\widetilde {\bf X}}^2}} + e A_0({\widetilde {\bf X}}) + H_F,
\label{ham}
\end{align}
where $H_F$ is the usual Hamiltonian for the free electromagnetic field:
\be
H_F = \int d^3 x \left[ \frac{1}{2} {\boldsymbol \Pi}^2 + \frac{1}{2}(\bn \times {\bf A})^2 + A_0 \bn \cdot {\boldsymbol \Pi} \right].
\label{fieldh}
\en
As is well known, there are two constraints: the primary constraint
\be
\Pi_0 = 0, 
\label{pi0}
\en
and the secondary constraint (which is the Gauss law)
\be
\bn \cdot {\boldsymbol \Pi}  + e \de({\bf x} - {\widetilde {\bf X}})= 0.
\label{gauss}
\en
The matter part of the Hamiltonian \eqref{ham} is not yet expressed in terms of the momenta. Using 
\be
\frac{m^2}{1 - \dot {\widetilde {\bf X}}(t)^2} = m^2 + \left({\widetilde {\bf P}} - e {\bf A}({\widetilde {\bf X}})\right)^2 ,
\en
we obtain
\be
H = \pm \sqrt{m^2 +  \left({\widetilde {\bf P}} - e {\bf A}({\widetilde {\bf X}})\right)^2 }  + e A_0({\widetilde{\bf X}}) + H_F,
\en
which involves the choice of a square root. 

Quantization in the Schr\"odinger picture (and dropping the tildes) leads to the following equations for $\Psi({\bf X},A,t)$:
\be
\ii \pa_t \Psi = \pm \sqrt{m^2 -  \left(\bn - \ii e {\bf A}({\bf X})\right)^2 } \Psi + e A_0({\bf X})\Psi + {\widehat H}_F\Psi,
\label{20}
\en
\be
\frac{\de \Psi}{\de A_0} =0, \qquad \ii {\boldsymbol \nabla} \cdot \frac{\de \Psi}{\de {\bf A}({\bf x})} - e \delta({\bf x} - {\bf X}) \Psi = 0,
\label{21}
\en
where
\be
{\widehat H}_F = \frac{1}{2}  \int d^3 x \left(-  \frac{\de^2 }{\de {\bf A}({\bf x})^2} + [{\boldsymbol \nabla} \times {\bf A}({\bf x})]^2 \right).
\label{fieldop}
\en
The equations \eqref{21} are the constraints \eqref{pi0} and \eqref{gauss} that are imposed as operator constraints. The square root could be eliminated by considering the square of \eqref{20} to obtain a Klein-Gordon-like equation. If $\lam$ rather than $t$ is taken as a common temporal parameter then this is indeed the equation that will be obtained, as we will now show.

To use $\lam$ as the temporal parameter, $1= \int d\lam \delta(\lam - \lam(t))$ is inserted in the field action. Defining ${\widetilde A}(\lam,{\bf x})= A(X^0(\lam),{\bf x})$, so that ${\widetilde A}'=\dot A X'^0$, the field action can be written as
\be
{\mathcal S}_F = \int dt L_F(A,\dot A) = \int d\lam L^*_F({\widetilde A},{\widetilde A}'),
\en
with 
\be
L^*_F({\widetilde A},{\widetilde A}') =  X'^0 L_F\left({\widetilde A},{\widetilde A}'/X'^0 \right).
\en
With ${\mathcal S}=\int d\lam L^*(X,X',{\widetilde A},{\widetilde A}')$, the conjugate momenta are
\be
{\widetilde \Pi}_\mu({\bf x}) = \frac{\de L^*}{\de {\widetilde A}'^\mu({\bf x})} =  \frac{\de L}{\de \dot A^\mu({\bf x})} = \Pi_\mu({\bf x}) ,
\en
\be
P_\mu = \frac{\pa L^*}{\pa X'^\mu} =-m \frac{X'_\mu}{\sqrt{{X'}^\nu X'_\nu }} - e {\widetilde A}_\mu({\bf X}) - \delta^0_\mu H_F({\widetilde A},{\widetilde \Pi}).
\en
So the expressions for the momenta for the field are just the same as before, cf.\ \eqref{fmom}. The particle momentum $P_0$ gets a contribution from field Lagrangian which is just minus the field Hamiltonian \eqref{fieldh}. 

The canonical Hamiltonian is
\be
H^* = P_\mu X'^\mu + \int d^3 x {\widetilde \Pi}_\mu({\bf x}){\widetilde A}'^\mu({\bf x}) - L^* = 0
\en
and is zero because of the reparameterization invariance of the action. There are two primary constraints: $\chi_1 = \Pi_0=0$ (as before) and 
\be
\chi_2 = \left[ P_\mu +  e {\widetilde A}_\mu({\bf X}) + \delta^0_\mu H_F({\widetilde A},{\widetilde \Pi}) \right]\left[ P^\mu +  e {\widetilde A}^\mu({\bf X}) + \delta^\mu_0 H_F({\widetilde A},{\widetilde \Pi}) \right] - m^2 =0.
\en
There is a secondary constraint which follows from the requirement that the Poisson bracket of the two primary constraints with the total Hamiltonian
\be
H^*_T = H^*  + \int d^3 x \lambda_1({\bf x}) \chi_1({\bf x})+ \lambda_2 \chi_2 
\en
vanishes, which results in $[\chi_1({\bf x}),\chi_2]_P = 0$ or
\be
\frac{\de H_F({\widetilde A},{\widetilde \Pi})}{ \de {\widetilde A}_0({\bf x})}  + e \de({\bf x} - {\bf X}) = 0,
\en
which amounts to the Gauss constraint \eqref{gauss}.

Quantization (again dropping the tildes) leads to the following equation for $\Psi(X,A)$:
\be
\left( \pa_\mu + \ii e A_\mu({\bf X}) + \ii \delta^0_\mu {\widehat H}_F \right) \left( \pa^\mu + \ii e  A^\mu({\bf X}) + \ii \delta^\mu_0 {\widehat H}_F \right)\Psi + m^2 \Psi = 0 ,
\label{25}
\en
with ${\widehat H}_F$ as before in \eqref{fieldop}, together with the constraints \eqref{21}. This Klein-Gordon-type equation is just the square of \eqref{20}, provided $X^0$ is identified with $t$.

So far we have dealt with just a single particle. How to extend the theory to many particles? The theory with the square root Hamiltonian \eqref{20} is directly generalized to many particles \eqref{20}, with now $\psi=\psi({\bf X}_1,\dots,{\bf X}_n, A,t)$:
\be
\ii \pa_t \Psi = \sum_k \left[ \sqrt{m^2 -  \left(\bn_k - \ii e {\bf A}({\bf X}_k)\right)^2 }  + e A_0({\bf X}_k) \right]\Psi + {\widehat H}_F\Psi,
\label{30}
\en
\be
\frac{\de \Psi}{\de A_0} =0, \qquad \ii {\boldsymbol \nabla} \cdot \frac{\de \Psi}{\de {\bf A}({\bf x})} - e \sum_k \delta({\bf x} - {\bf X}_k) \Psi = 0.
\label{31}
\en
We have chosen the positive square root Hamiltonian for each particle. The Klein-Gordon-type equation \eqref{25} is not that straightforwardly generalizable to many particles because it is not of the Schr\"odinger form like \eqref{20}. One option is to run through the quantization procedure again for many particles, but this seems rather complicated. Another option is to recast the Klein-Gordon-type equation \eqref{25} in a Schr\"odinger form, which could be done using the Kemmer formulation \cite{kemmer39} (which actually concerns a Dirac-like equation for spin-0), and which at least in the case of external field is directly extendable to many particles \cite{struyve03b}. Yet another option is to use the multi-time picture where the wave function depends on the time-component of each of the particles \cite{petrat14}. We will not pursue this further here.

\section{Gravity}
The analysis of gravity proceeds completely analogously as that of electromagnetism. The action for a classical point-particle coupled to gravity is 
\be
{\mathcal S} =  {\mathcal S}_M + {\mathcal S}_G ,
\en
with
\be
{\mathcal S}_M = - m \int d\lambda \sqrt{g_{\mu \nu}(X(\lam)){X'}^\mu(\lam) X'^\nu (\lam)} 
\en
and
\be
{\mathcal S}_G = -\frac{1}{\ka} \int d^4x \sqrt{-g} R ,
\en
with $\ka = 16\pi G$, is the Einstein-Hilbert action. 

Before considering the Hamiltonian picture in terms of the temporal parameter $\lam$, we recall the usual Hamiltonian formulation for the Einstein-Hilbert action in terms of $t$ (following the conventions of \cite{carlip19}). It is supposed that space-time can be foliated in terms of space-like hypersurfaces such that the space-time manifold is diffeomorphic to ${\mathbb R} \times \Sigma$, with $\Sigma$ a 3-surface. Coordinates $x^\mu=(t,{\bf x})$ are chosen such that the time coordinate $t$ labels the leaves of the foliation and ${\bf x}$ are the coordinates on $\Sigma$. In terms of these coordinates, the metric and its inverse are written as
\be
g_{\mu \nu}=
\begin{pmatrix}
N^2 - N_i N^i & -N_i \\
-N_i & - h_{ij}
\end{pmatrix} \,,
\qquad
g^{\mu \nu}=
\begin{pmatrix}
\frac{1}{N^2} & \frac{-N^i}{N^2} \\
\frac{-N^i}{N^2} &   \frac{N^iN^j}{N^2}- h^{ij}
\end{pmatrix} \,,
\label{199}
\en
where $N$ and $N^i$ are respectively the lapse and the shift vector, and $h_{ij}$ is the Riemannian metric on $\Sigma$, with $h$ its determinant. Spatial indices of $h_{ij}$ and $N^i$ and the corresponding momenta are raised and lowered with this spatial metric. 

With $L_G$ the gravitational Lagrangian, the canonically conjugate momenta are 
\be
\pi^{ij} = \frac{\delta L_G }{ \delta {\dot h}_{ij}} = - \frac{1}{\kappa} \sqrt{h} (K^{ij} - K h^{ij})
\en
(which is a tensor density of weight $-1$), with 
\be
K_{ij} = \frac{1}{2N}({ D _i N_j + D _j N_i - \dot{h}}_{ij}) ,\qquad K = K_{ij}h^{ij},
\en
the extrinsic curvature, and
\be
\pi = \frac{\delta L_G }{ \delta N} = 0, \qquad \pi_i = \frac{\delta L_G }{ \delta N^i} =0.
\label{momcon}
\en
The Hamiltonian is
\be
H_G = \int_\Sigma d^3 x \left(N  {\mathcal H} + N^i  {\mathcal H}_i  \right),
\en
with
\be
 {\mathcal H} = \ka G_{ijkl} \pi^{ij} \pi^{kl} + {\mathcal V}, \qquad   {\mathcal H}_{i} =  -2 h_{ik} D_j \pi^{jk}  ,
\label{hamdif}
\en
where $G_{ijkl} = (h_{ij} h_{jl} + h_{il} h_{jk} -  h_{ij} h_{kl})/2\sqrt{h}$ is the DeWitt metric, $D_i$ is the covariant derivative corresponding to $h_{ij}$, and ${\mathcal V} = -\sqrt{h} R^{(3)}/\ka$ is the gravitational potential density. Apart from the primary constraints \eqref{momcon}, there are also the secondary constraints 
\be
{\mathcal H} = 0, \qquad {\mathcal H}_i = 0.
\en

In order to use the temporal variable $\lam$, we introduce ${\widetilde g}_{\mu \nu}(\lam, {\bf x}) = g_{\mu \nu}(X^0(\lam), {\bf x})$ and similarly for other variables. The total action is
\be
{\mathcal S} = \int d\lam L^*(X,X',{\widetilde g},{\widetilde g}').
\en
The momenta for the metric are the same as before, i.e.,
\be
{\widetilde \pi}^{ij} = \frac{\de L^*}{\de {\widetilde h}'_{ij}}= \frac{\de L_G}{\de \dot h_{ij}} =  \pi^{ij} , \quad {\widetilde \pi} = \frac{\delta L^* }{ \delta {\widetilde N}} = 0, \quad {\widetilde \pi}_i = \frac{\delta L^* }{ \delta {\widetilde N}^i} =0,
\en
and the momentum for the particle becomes (indices of $X^\mu$ and $P_\mu$ are raised and lowered with the metric ${\widetilde g}_{\mu \nu}({\bf X})$)
\be
P_\mu = \frac{\pa L^*}{\pa X'^\mu}=  -m \frac{X'_\mu}{\sqrt{{X'}^\nu X'_\nu }}  - \delta^0_\mu {\widetilde H}_G,
\en
where 
\be
{\widetilde H}_G = H_G({\widetilde N},{\widetilde N}^i,{\widetilde g}_{ij},{\widetilde \pi}^{ij}).
\label{tildeh}
\en
 The canonical Hamiltonian is zero, i.e.,
\be
H^* = P_\mu X'^\mu + \int_\Sigma d^3 x {\widetilde \pi}^{ij}({\bf x}){\widetilde g}'_{ij}({\bf x}) - L^* = 0.
\en
There are three primary constraints: 
\be
{\widetilde \pi}=0, \qquad {\widetilde \pi}_i = 0,
\label{498}
\en
\be
\chi = {\widetilde g}^{\mu \nu}({\bf X})\left[ P_\mu +  \delta^0_\mu {\widetilde H}_G \right]\left[ P_\nu + \delta^0_\nu {\widetilde H}_G\right] - m^2=0.
\label{499}
\en
There are two secondary constraints corresponding to $\delta \chi/\delta {\widetilde N} = 0$ and $\delta \chi/\delta {\widetilde N}^i = 0$, resulting in
\be
{\widetilde {\mathcal H}}({\bf x}) = \de({\bf x} - {\bf X})\left({\widetilde N}({\bf X}) p^0 + \frac{1}{{\widetilde N}({\bf X})} {\widetilde H}_G\right) 
\label{500}
\en
and 
\be
{\widetilde {\mathcal H}}_i({\bf x}) =  \de({\bf x} - {\bf X})p_i,
\label{501}
\en
where (as in \eqref{tildeh}) the tilde on ${\widetilde {\mathcal H}}$ and ${\widetilde {\mathcal H}}_i$ denotes the functions \eqref{hamdif} evaluated for fields with tildes. There are no further constraints. 

By multiplying \eqref{500} with ${\widetilde N}({\bf x})$ and \eqref{501} with ${\widetilde N}^i({\bf x})$, integrating over all space, and using $p^0=(p_0 - {\widetilde N}^i({\bf X})p_i)/{\widetilde N}({\bf X})^2$, it follows that 
\be
p_0=0. 
\label{502}
\en
Hence, another interesting consequence is that
\be
{\widetilde N}({\bf x}){\widetilde {\mathcal H}}({\bf x}) + {\widetilde N}^i({\bf x}){\widetilde {\mathcal H}}_i({\bf x}) = \de({\bf x} - {\bf X}) {\widetilde H}_G.
\en

Summarizing, the Hamiltonian dynamics is completely determined by the constraints \eqref{498}-\eqref{501}. Using \eqref{502}, they can be simplified to (dropping the tildes):
\be
\pi=0, \qquad \pi_i = 0,
\label{504}
\en
\be
\left[H_G - N^i({\bf X}) p_i\right]^2 - N({\bf X})^2\left[h^{ij}({\bf X})p_ip_j + m^2\right] = 0,
\label{505}
\en
\be
{\mathcal H}({\bf x}) = \de({\bf x} - {\bf X})\frac{1}{N({\bf X})} \left[- N^i({\bf X})p_i + H_G \right],
\label{506}
\en
\be
{\mathcal H}_i({\bf x}) = \de({\bf x} - {\bf X}) p_i .
\label{507}
\en

Quantization in the Schr\"odinger picture leads to the following equations for $\Psi({\bf X},h_{ij})$:{\footnote{We have chosen the Laplace-Beltrami operator ordering for the particle but not for gravity \cite{christodoulakis86}.}}
\be
\left[\ii  N^i({\bf X})\nabla_i + {\widehat H}_G   \right]^2 \Psi - N({\bf X})^2\left(-\nabla^2 + m^2\right)\Psi  = 0,
\label{505.1}
\en
\be
{\widehat {\mathcal H}}({\bf x}) \Psi = \de({\bf x} - {\bf X})\frac{1}{N({\bf X})} \left[\ii N^i({\bf X})\nabla_i + {\widehat H}_G \right]\Psi ,
\label{506.1}
\en
\be
{\widehat {\mathcal H}}_i({\bf x}) \Psi = - \ii \de({\bf x} - {\bf X}) \nabla_i \Psi ,
\label{507.1}
\en
where $\nabla_i$ is the covariant derivative with respect to the metric $h_{ij}({\bf X})$, with $\nabla_i \psi = \pa_i \psi$, $\nabla^2 = \na_i\na^i$ is the Laplacian, and
\be
{\widehat H}_G = \int_\Sigma d^3 x \left(N  {\widehat {\mathcal H}} + N^i  {\widehat {\mathcal H}}_i  \right),
\en
\be
{\widehat {\mathcal H}} =  -\ka G_{ijkl} \frac{\delta^2}{\delta h_{ij}\delta h_{kl}}  + {\mathcal V}(h,\phi), \qquad {\widehat {\mathcal H}}_i =  2\ii h_{ik}D_j\frac{\delta }{\delta h_{jk}}.
\en
The wave function $\Psi$ does not depend on $N$, $N^i$ and $X^0$, because of the operator constraints following from \eqref{502} and \eqref{504}. The latter means that the wave functional does not depend on time as is familiar in Wheeler-DeWitt quantization. 

The wave functional does not depend on $N$ and $N^i$, but they still appear in the wave equations. Choosing $N=1$ and $N^i=0$, results in
\be
\left( {\widehat H}_G^2  + \nabla^2 - m^2 \right)\Psi  = 0,
\label{510}
\en
\be
{\widehat {\mathcal H}}({\bf x}) \Psi = \de({\bf x} - {\bf X}) {\widehat H}_G\Psi ,
\label{511}
\en
\be
{\widehat {\mathcal H}}_i({\bf x}) \Psi = - \ii \de({\bf x} - {\bf X}) \nabla_i \Psi .
\label{512}
\en

Applying the quantization recipe using $t$ as temporal coordinate leads to the following quantum theory \cite{rovelli91a,rovelli91b}:
\be
{\widehat {\mathcal H}}({\bf x}) \Psi = \pm \de({\bf x} - {\bf X})  \sqrt{-\nabla^2 + m^2 } \Psi, \qquad {\widehat {\mathcal H}}_i({\bf x}) \Psi = - \ii \de({\bf x} - {\bf X}) \nabla_i \Psi,
\label{hamt}
\en
with $\Psi({\bf X},h_{ij})$. This theory also follows from \eqref{504}-\eqref{507} if the square root is taken in \eqref{505}. As in the case of electromagnetism, the extension to many particles is straightforward, leading to
\be
{\widehat {\mathcal H}}({\bf x}) \Psi = \sum_k  \de({\bf x} - {\bf X}_k) \sqrt{-\nabla^2_k + m^2 } \Psi, \qquad {\widehat {\mathcal H}}_i({\bf x}) \Psi = - \ii \sum_k \de({\bf x} - {\bf X}_k) \nabla_{ki} \Psi,
\label{hamt2}
\en
with $k$ the particle label, $\Psi({\bf X}_1,\dots,{\bf X}_n ,h_{ij})$. The extension of the Klein-Gordon form \eqref{510}-\eqref{512} to many particles requires a bit more effort (though apparently less than in the case of electromagnetism due to absence of the time derivative). The natural generalization seems to be 
\be
\left( {\widehat O}^2_k  + \nabla^2_k - m^2 \right)\Psi  = 0,
\label{510.1}
\en
\be
{\widehat {\mathcal H}}({\bf x}) \Psi = \sum_k \de({\bf x} - {\bf X}_k) {\widehat O}_k\Psi ,
\label{511.1}
\en
\be
{\widehat {\mathcal H}}_i({\bf x}) \Psi = - \ii \sum_k \de({\bf x} - {\bf X}_k) \nabla_{ki} \Psi .
\label{512.1}
\en
The operators ${\widehat O}_k$ can be determined almost completely by self-consistency. Namely, for configurations $({\bf X}_1,\dots,{\bf X}_N)$ with all the ${\bf X}_k$ different, we can integrate \eqref{511.1} over a test function $f({\bf x})$ such that $f({\bf X}_k)=1$ and  $f({\bf X}_l)=0$ for $l \neq k$, to obtain $ {\widehat O}_k \Psi= \int d^3x f({\bf x}){\widehat {\mathcal H}}({\bf x}) \Psi $. If not all the ${\bf X}_k$ are different, the action of the ${\widehat O}_k$ could be defined similarly by requiring symmetry. For example, for a configuration with ${\bf X}_k$ = ${\bf X}_{l}$, we can require that ${\widehat O}_k \Psi = {\widehat O}_l\Psi$. This generalization is consistent with \eqref{hamt2}, in the sense that \eqref{hamt2} follows by taking particular square roots in \eqref{510.1}.

The Wheeler-DeWitt quantization of the relativistic particle was considered before by Pav{\v s}i{\v c} \cite{pavsic11}. However, in the quantization procedure no common temporal parameter was used. In the definition of the particle momentum the parameter $\lam$ was used, while in the definition of the field momenta the parameter $t$ was used. As such the quantization deviates from the usual recipe and the resulting quantum theory does not agree with the one presented here. However, the equations obtained by Pav{\v s}i{\v c} follow from ours when interpreted in a partial Heisenberg picture (and as such is reminiscent of the Dirac-Fock-Podolsky formulation of a particle interacting with an electromagnetic field \cite{dirac32}). That is, defining 
\be
\Phi(X,h_{ij}) = \ee^{\ii {\widehat H}_G X^0 } \Psi({\bf X},h_{ij}) ,\qquad h_{ij}(X) = \ee^{\ii {\widehat H}_G X^0 } h_{ij}({\bf X}) \ee^{-\ii {\widehat H}_G X^0 } ,
\en
\be
{\widehat {\mathcal H}}({\bf x},X^0) = \ee^{\ii {\widehat H}_G X^0 } {\widehat {\mathcal H}}({\bf x}) \ee^{-\ii {\widehat H}_G X^0 },\qquad {\widehat {\mathcal H}}_i({\bf x},X^0) = \ee^{\ii {\widehat H}_G X^0 } {\widehat {\mathcal H}}_i({\bf x}) \ee^{-\ii {\widehat H}_G X^0 },
\en
the equations \eqref{510}-\eqref{512} reduce to
\be
\left( \pa^2_0  - {\widetilde \nabla}^2 + m^2 \right)\Phi  = 0,
\label{553}
\en
\be
{\widehat {\mathcal H}}({\bf x},X^0) \Phi = -\ii \de({\bf x} - {\bf X}) \pa_0 \Phi ,
\label{554}
\en
\be
{\widehat {\mathcal H}}_i({\bf x},X^0) \Phi = - \ii \de({\bf x} - {\bf X}) {\widetilde \nabla}_i \Phi ,
\label{555}
\en
where ${\widetilde \nabla}$ and ${\widetilde \nabla}^2$ concern the covariant derivative with respect to the metric $h_{ij}(X)$. (Pav{\v s}i{\v c} also chooses a different operator ordering, but that difference is not essential.) For Pav{\v s}i{\v c} the appearance of the time derivative in \eqref{554} is interesting the light of the problem of time. However, the proper quantum theory is time-independent.

\section{Acknowledgments}
This work is supported by the Research Foundation Flanders (Fonds Wetenschappelijk Onderzoek, FWO), Grant No. G066918N. It is a pleasure to thank Christian Maes and Kasper Meerts for useful discussions.


\begin{thebibliography}{10}

\bibitem{kiefer04}
{C.\ Kiefer, {\em Quantum Gravity}, International Series of Monographs on
  Physics {\bf 124}, Clarendon Press, Oxford (2004).}

\bibitem{kuchar92}
{K.V.\ Kucha{\u{r}}, ``Time and interpretations of quantum gravity'', in {\em
  Proceedings of the 4th Canadian Conference on General Relativity and
  Relativistic Astrophysics}, eds.\ G.\ Kunstatter, D.\ Vincent and J.\
  Williams, World Scientific, Singapore (1992), reprinted in {\em Int.\ J.\
  Mod.\ Phys.\ D} {\bf 20}, 3-86 (2011).}

\bibitem{rovelli04}
{C.\ Rovelli, {\em Quantum Gravity}, Cambridge University Press, Cambridge
  (2004).}

\bibitem{pinto-neto19}
{N.\ Pinto-Neto and W.\ Struyve, ``Bohmian quantum gravity and cosmology'', in
  {\em Applied Bohmian Mechanics: From Nanoscale Systems to Cosmology}, 2nd
  edition, eds.\ X.\ Oriols and J.\ Mompart, 607-656 (2019) and
  arXiv:1801.03353 [gr-qc].}

\bibitem{rovelli91a}
{C.\ Rovelli, ``What is observable in classical and quantum gravity?'', {\em
  Class.\ Quantum Grav.}\ {\bf 8}, 297-316 (1991).}

\bibitem{rovelli91b}
{C.\ Rovelli, ``Quantum reference systems'', {\em Class.\ Quantum Grav.}\ {\bf
  8}, 317-331 (1991).}

\bibitem{pavsic11}
{M.\ Pav{\v s}i{\v c}, ``Klein--Gordon--Wheeler--DeWitt--Schr\"odinger
  equation'', {\em Phys.\ Lett.\ B} {\bf 703}, 614-619 (2011) and
  arXiv:1106.2017 [gr-qc].}

\bibitem{pavsic20}
{M.\ Pav{\v s}i{\v c}, ``A novel approach to quantum gravity in the presence of
  matter without the problem of time'', {\em Int.\ J.\ Mod.\ Phys.\ A} {\bf
  35}, 2050015 (2020) and arXiv:1906.03987 [physics.gen-ph].}

\bibitem{brown95}
{J.D.\ Brown and K.V.\ Kucha{\v{r}}, ``Dust as a standard of space and time in
  canonical quantum gravity'', {\em Phys.\ Rev.\ D} {\bf 51}, 5600 (1995) and
  arXiv:gr-qc/9409001.}

\bibitem{dirac64}
{P.A.M.\ Dirac, {\em Lectures On quantum Mechanics}, Belfer Graduate School of
  Science, Yeshiva University, New York (1964), reprinted by Dover
  Publications, New York (2001).}

\bibitem{hanson76}
{A.\ Hanson, T.\ Regge and C.\ Teitelboim, {\em Constrained Hamiltonian
  Systems}, Accademia Nazionale Dei Lincei, Roma (1976).}

\bibitem{sundermeyer82}
{K.\ Sundermeyer, {\em Constrained Dynamics}, Lecture Notes in Physics 169,
  Springer-Verlag, Berlin (1982).}

\bibitem{gitman90}
{D.M.\ Gitman and I.V.\ Tyutin, {\em Quantization of Fields with Constraints},
  Springer-Verlag, Berlin Heidelberg (1990).}

\bibitem{henneaux91}
{M.\ Henneaux and C.\ Teitelboim, {\em Quantization of Gauge Systems},
  Princeton University Press, New Jersey (1991).}

\bibitem{kemmer39}
{N.\ Kemmer, ``The particle aspect of meson theory'', {\em Proc.\ R.\ Soc.\ A}
  {\bf 173}, 91-116 (1939).}

\bibitem{struyve03b}
{W.\ Struyve, W.\ De Baere, J.\ De Neve and S.\ De Weirdt, ``On the uniqueness
  of paths for spin-0 and spin-1 quantum mechanics'', {\em Phys.\ Lett.\ A}
  {\bf 322}, 84-95 (2004) and arXiv:quant-ph/0311098.}

\bibitem{petrat14}
{S.\ Petrat and R.\ Tumulka, ``Multi-time equations, classical and quantum'',
  {\em Proc.\ R.\ Soc.\ A} {\bf 470}, 20130632 (2014) and arXiv:1309.1103
  [quant-ph].}

\bibitem{carlip19}
{S.\ Carlip, {\em General Relativity}, Oxford University Press, Oxford (2019).}

\bibitem{christodoulakis86}
{T.\ Christodoulakis and J.\ Zanelli, ``Operator Ordering in Quantum Mechanics
  and Quantum Gravity'', {\em Nuovo Cimento B} {\bf 93}, 1-21 (1986).}

\bibitem{dirac32}
{P.A.M.\ Dirac, V.A.\ Fock and B.\ Podolsky, ``On quantum electrodynamics'',
  {\em Physikalische Zeitschrift der Sowjetunion} {\bf 2}, 468–479 (1932);
  reprinted in `Selected Papers on Quantum Electrodynamics'', ed.\ J.\
  Schwinger, Dover, New York (1958).}

\end{thebibliography}
\end{document}